\documentclass[twocolumn,showpacs,preprintnumbers,amsmath,amssymb]{revtex4}

\usepackage{graphicx}
\usepackage{subfigure}

\begin{document}



\title{Time delay between photoemission from the 2p and 2s subshells of Neon
}


\author{L. R. Moore}
\affiliation{Centre for Theoretical Atomic, Molecular and Optical Physics, Queen's University Belfast,
Belfast BT7~1NN, UK}

\author{M. A. Lysaght}
\altaffiliation{Present address: Department of Physics and Astronomy, The Open University, Walton Hall,
Milton Keynes MK7~6AA, UK}
\affiliation{Centre for Theoretical Atomic, Molecular and Optical Physics, Queen's University Belfast,
Belfast BT7~1NN, UK}

\author{J. S. Parker}
\affiliation{Centre for Theoretical Atomic, Molecular and Optical Physics, Queen's University Belfast,
Belfast BT7~1NN, UK}

\author{H. W. van der Hart}
\email[h.vanderhart@qub.ac.uk]{}
\affiliation{Centre for Theoretical Atomic, Molecular and Optical Physics, Queen's University Belfast,
Belfast BT7~1NN, UK}

\author{K. T. Taylor}
\affiliation{Centre for Theoretical Atomic, Molecular and Optical Physics, Queen's University Belfast,
Belfast BT7~1NN, UK}


\begin{abstract}
The {\it R}-Matrix incorporating Time (RMT) method is a new 
method for solving the time-dependent Schr\"{o}dinger equation for multi-electron 
atomic systems exposed to intense short-pulse laser light. We have
employed the RMT method to investigate the time delay in the photoemission 
of an electron liberated from a 2p orbital in a neon atom with respect to 
one released from a 2s orbital following absorption of an attosecond XUV pulse.
Time delays due to XUV pulses in the range 76-105~eV are presented. 
For an XUV pulse at the experimentally relevant 105.2~eV, we calculate 
the time delay to be $10.2 \pm 1.3$ attoseconds, somewhat larger than 
estimated by other theoretical calculations, but still a factor two smaller 
than experiment. We repeated the calculation for  a photon energy of 89.8~eV
with a larger basis set capable of 
modelling correlated-electron dynamics within the neon atom and the residual 
Ne$^+$ ion. A time delay of $14.5 \pm 1.5$~attoseconds was observed,
compared to a $16.7 \pm 1.5$~attosecond result using a single-configuration
representation of the residual Ne$^+$ ion.
\end{abstract}


\maketitle

\normalsize

One of the goals of attosecond science is to provide insights into the behaviour of atomic electrons by imaging and controlling electronic motion using intense laser beams \cite{corkum2007}.  Recent advances in attosecond technology allow the delivery of light pulses with high intensity and with durations in the attosecond (as) range \cite{kling2008,sansone2006}.  Such revolutionary laser technology is enabling time-resolved measurements of correlated electron dynamics in atomic systems.  One such tool for achieving time resolution on the sub-100~as time scale is attosecond streaking \cite{goulielmakis2004,kienberger2004}.

Attosecond streaking is based on a pump-probe experiment, in which an extreme ultraviolet (XUV) pulse of duration a few hundred attoseconds is used as the pump and a low intensity phase-controlled few-cycle infrared (IR) pulse as the probe.  The time delay between the two fields is varied but is such that the XUV and IR pulses overlap in time.  The XUV pulse causes the emission of a photoelectron.  Upon ejection, this electron is accelerated by the IR field.  Its final energy and momentum depend on the value of the IR vector potential at the moment of its escape.  Thus information on the time of ejection of the electron is embedded in its final escape energy.

Recently an attosecond streaking experiment investigated the time delay in photoemission of electrons liberated from the 2p orbitals of neon atoms with respect to those released from the 2s orbital by the same XUV light pulse \cite{Schultze2010}.  This time delay was measured to be $21 \pm 5$~as, suggesting a small delay time between the formation of electron wavepackets originating from the two different valence sub-shells.  Theoretical calculations suggest that the delay is substantially smaller than the measured value.  An independent electron model with a correlation correction calculated the delay to be 6.4~as \cite{Schultze2010}.  An independent method using Hartree Fock (HF) phase derivatives together with the random phase approximation with exchange (RPAE) correction for correlation calculated the delay to be 8.4~as \cite{Kheifets2010}.

The question arises as to whether the discrepancies between theory and experiment are related to the way in which the many-electron correlation effects are handled in the theoretical descriptions of the laser-atom interactions.  Both theoretical methods mentioned above added a correlation correction {\em ad hoc}.  It is possible that a solution of the full time-dependent Schr\"{o}dinger equation (TDSE) taking multi-electron correlation effects directly into account could modify the theoretical results.

We have recently developed a new {\em ab initio} method for solving directly and accurately the TDSE describing the detailed response of multi-electron atoms and ions to short intense pulses of laser light: the $R$-Matrix incorporating Time (RMT) method \cite{Moore2011,Lysaght2011}.  The RMT method utilizes the powerful {\it R}-matrix theory {\it division-of-space} concept \cite{burke1993} to split
the position space occupied by the atomic electrons into two regions: a multi-electron inner region and an outer region in which one outer electron has become separated from the other electrons.
In the multi-electron inner region, electron-electron interactions are fully described and multi-electron atom-laser Hamiltonian matrix elements are calculated explicitly.  The multi-electron wavefunction in this region is constructed from basis functions which have the form of a close-coupling expansion with pseudostates \cite{burke1993}.  In the outer region only one electron is present and the electron there, besides experiencing the laser field directly, is aware of the remainder of the atomic system only via long-range multi-pole interactions.  This effective one-electron problem is solved using state-of-the-art grid-based technology \cite{smyth1998}.  In both spatial regions the TDSE is integrated using high-order explicit time propagator methods \cite{arnoldi1951}.  A central concept of the RMT method, namely the matching of a finite-difference representation in one region with a basis set representation in the other, was first developed using the hydrogen atom as a testing ground \cite{nikolopoulos2008}.

In this communication, we present an application of RMT to the time delay between photoemission from the 2s and 2p sub-shells of a neon atom.  Laser pulse parameters are chosen to closely resemble those used in the experiment \cite{Schultze2010}.  The 800~nm IR field has intensity $10^{11}$~W/cm$^2$ and is linearly polarized in the $z$-direction.  It is a three-cycle pulse with a sin$^2$ profile.  The XUV pulse has a central photon energy of 105.2~eV and peak intensity $10^{13}$~W/cm$^2$.  It is also linearly polarized in the $z$-direction and has a Gaussian profile with a full-width at half-maximum (FWHM) in intensity of 270~as.

The RMT inner region has a radius of 20~a.u..  The multi-electron wavefunction in this region is expanded on a basis of field-free $R$-matrix eigenfunctions.  The $R$-matrix basis used is one developed for single-photon ionization of Ne \cite{taylor1975}.  Our calculations include both the 1s$^2$2s$^2$2p$^5$~$^2{\rm P}^o$ ground state and the 1s$^2$2s2p$^6$~$^2{\rm S}^e$ excited state of Ne$^+$, so that both emission of a 2p electron and of a 2s electron are accounted for in the same calculation.  In the present calculations both ionic states are represented by single configurations using HF orbitals for the $^2{\rm P}^o$ ground state of the ion.
The description of Ne includes all 1s$^2$2s$^2$2p$^5 \epsilon l$ and all 1s$^2$2s2p$^6 \epsilon l$ channels up to a given total angular momentum $L = L_{\rm max}$ where $L_{\rm max}=9$.  The ionization potential of the ground state is set to the experimental value \cite{saloman2004}.
The RMT outer region radial wavefunction is discretized on a finite difference grid extending to 6600~a.u. with a grid spacing of 0.2~a.u..

In the time integration we use a time step of 0.0125~a.u..  Typically we let the calculation run for a further 15000 time steps after the IR laser pulse has passed.  This allows the ejected electronic wavepackets to propagate to regions far from the nucleus where the Coulombic field of the residual ion is weak.  We have checked all parameter settings (including the radius of the inner region) used in the code to ensure that results are converged.
Similarly we have checked the laser parameters to ensure they have no effect on results.
The only exception we have found is that reducing the duration of the XUV pulse can impact on the time delay: at extremely short durations the Fourier transform of the pulse profile itself can have substantial side wings, effectively introducing additional XUV frequencies into the calculation.

Upon completion of the time propagation, the outer-electron wavefunction needs to be decoupled from the full multi-electron wavefunction \cite{vanderhart2008}.  This outer-electron wavefunction is then transformed to momentum space under the assumption that the long-range Coulomb potential is negligible, a valid assumption since the ejected electronic wavepackets are propagated to regions sufficiently far from the nucleus so that a field-free spherical wave transformation is satisfactory.

\begin{figure}[]
\begin{center}
\subfigure[]
{\includegraphics[clip=true,width=5.0cm]{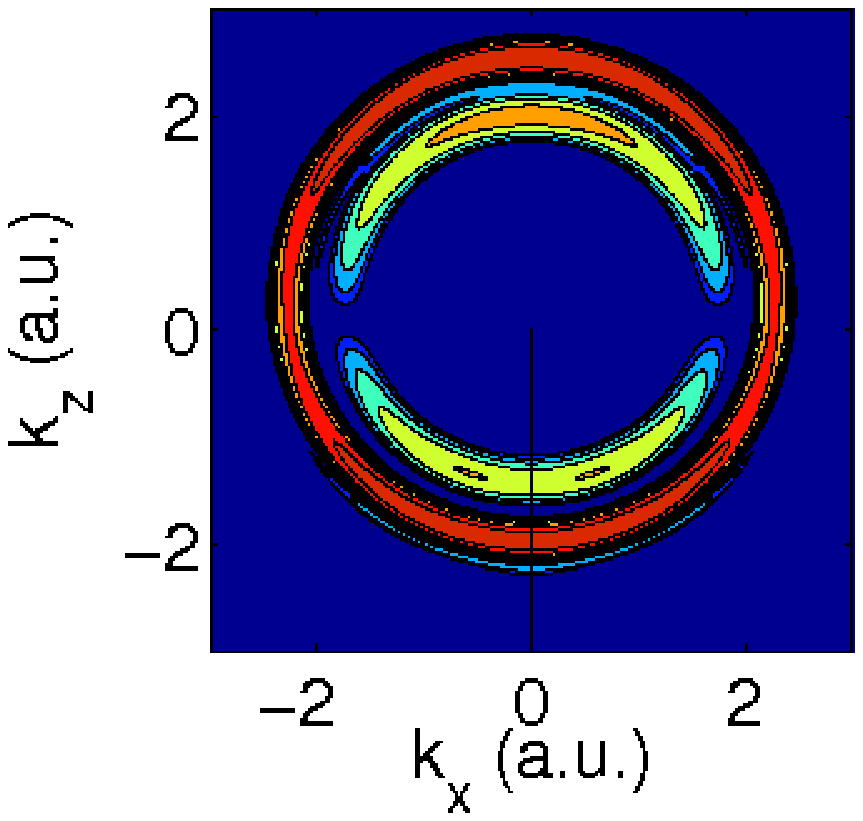}
\label{fig:kxkza}
}
\subfigure[ ]
{\includegraphics[clip=true,width=5.0cm]{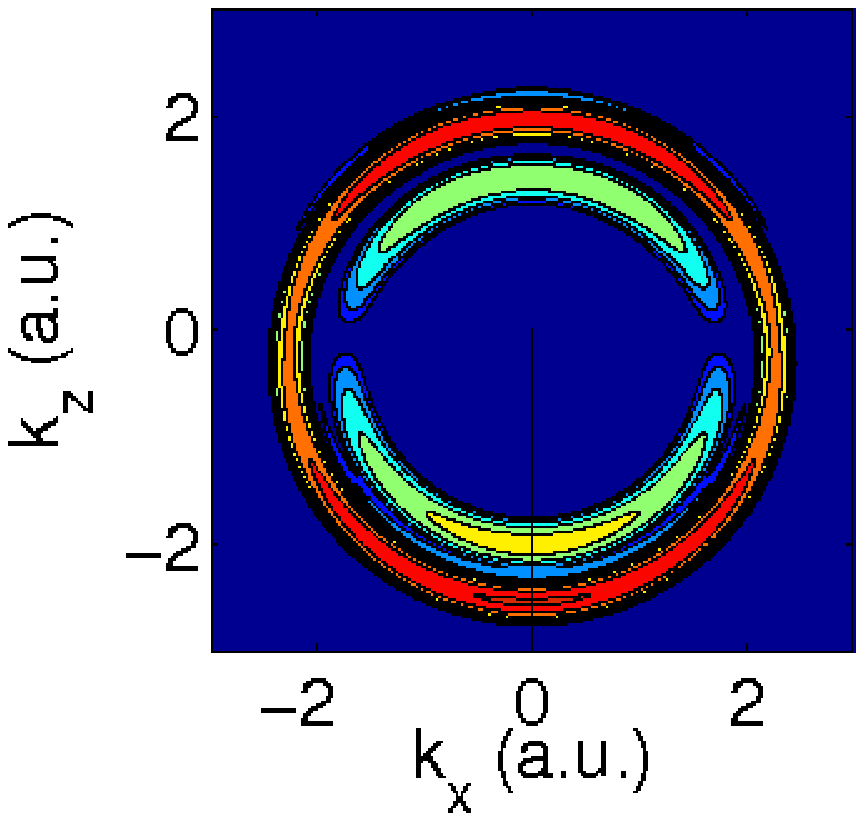}
\label{fig:kxkzb}
}
\caption{\label{fig1} (Colour online) Momentum spectra in the $x-z$ plane obtained for the two cases of the
 XUV field delayed, relative to the IR field, by (a) 1.25 ${\rm T}_{\rm IR}$
 and (b) 1.75 ${\rm T}_{\rm IR}$.
  In this example the intensity
 of the IR field is increased to $10^{13}$~W/cm$^2$.}%
\label{fig:kxkz}
\end{center}
\end{figure}

Figure~\ref{fig:kxkz} displays momentum spectra in the $x-z$ plane of the outgoing electron for the two cases of the XUV field delayed, relative to the IR field, by (a) 1.25 ${\rm T}_{\rm IR}$ and (b) 1.75 ${\rm T}_{\rm IR}$, where ${\rm T}_{\rm IR}$ is the IR laser field period.  In Fig.~\ref{fig:kxkza} the peak intensity of the XUV field coincides with a minimum in the IR vector potential, and in Fig.~\ref{fig:kxkzb} with a maximum.  Because the drift momentum, $P_z$, of the outgoing electron is equal to the charge on the electron (-1) times the value of the IR vector potential at the time of emission, the momentum spectrum is shifted upwards in Fig.~\ref{fig:kxkza} and downwards in Fig.~\ref{fig:kxkzb} relative to a position centred on the origin.  Visible in both plots of Fig.~\ref{fig:kxkz} are two rings.  The outer (inner) ring represents the component of the outgoing wavepacket arising from ionization of the 2p (2s) sub-shell.

\begin{figure}
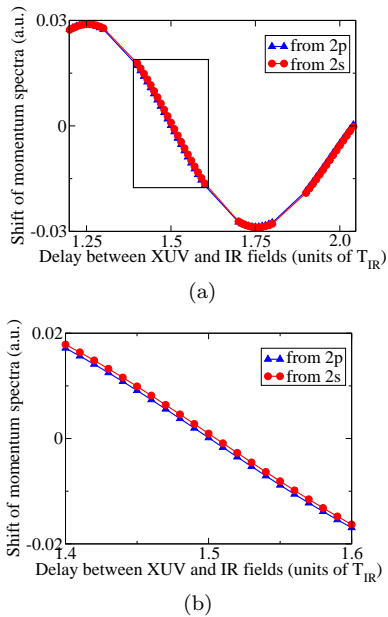

\begin{center}
\subfigure[]
{\includegraphics[clip=true,width=5.00cm]{figure2a.eps}
\label{fig:timedelaya}
}
\subfigure[ ]
{\includegraphics[clip=true,width=5.00cm]{figure2b.eps}
\label{fig:timedelayb}
}
\label{fig:timedelay}
\caption{\label{fig2} (Colour online) The shifts of the momentum distributions of electrons ejected from the 2s and from the 2p sub-shells as a function of the
 delay between the XUV and IR fields.
 Figure \ref{fig:timedelayb}: Magnification of the region enclosed by the black box in Fig. \ref{fig:timedelaya}.}%
\end{center}
\end{figure}

Further calculations are carried out at different delays between the XUV and IR fields.  For each delay, the shifts of both the inner and the outer rings of the momentum spectrum are calculated.  These shifts correspond to the additional momentum $P_z$, imparted by the IR field, gained by electrons emitted from the 2s and from the 2p sub-shells respectively.  Figure \ref{fig:timedelaya} plots these shifts as a function of delay between the XUV and IR fields.
As is evident from Fig.~\ref{fig:timedelayb}, the shifts of the momentum distributions are not identical for electrons emitted from the two different valence shells: they exhibit a small temporal shift with respect to one another.
It is as though the electrons emitted from the two valence shells experience a slightly different IR vector potential at the moment of release, thus suggesting a time delay in photoemission as observed experimentally.  By fitting a polynomial through the data points shown in Fig.~\ref{fig:timedelayb}, we calculate that an electron emitted from the 2p orbital is delayed by 10.2~as relative to one released from the 2s orbital.  The use of fitting procedures also provides a measure of the accuracy with which the delay can be extracted from our calculations.  Furthermore we have repeated the calculations using a variety of XUV pulse profiles.  Through these additional investigations, we estimate the standard deviation in our calculated time delay to be $1.3$~as.

Table~\ref{tab:table1} presents a comparison of our calculated value for the time delay with other values calculated both numerically and experimentally.  The RMT result is slightly higher than the other theoretical results, but it too lies substantially below the experimental result.

\begin{table}
\caption{\label{tab:table1} Comparison of time delays between photoemission from the 2p and 2s subshells of Neon following absorption of an attosecond XUV pulse of photon energy 105 to 106~eV. }
\begin{ruledtabular}
\begin{tabular}{p{0.27\linewidth}p{0.47\linewidth}p{0.26\linewidth}}
\hline
Group & Method & Delay (as)  \\
\hline
Schultze {\it et al} \cite{Schultze2010} &   Experiment &   $21 \pm 5$  \\
 & & \\
Schultze {\it et al} \cite{Schultze2010} &   Independent electron model & 4.0 \\
                     &   Correlation correction   & 2.4 \\
                     &   Total         & 6.4 \\
 & & \\
Kheifets {\it et al} \cite{Kheifets2010} &  HF phase derivatives &  6.2 \\
                     & RPAE correction   & 2.2 \\
                     & Total & 8.4 \\
 & & \\
Present work   & RMT  &  $10.2 \pm 1.3$\\
\end{tabular}
\end{ruledtabular}
\end{table}

\begin{figure}[b]
\begin{center}
\includegraphics[clip=true,width=6.25cm]{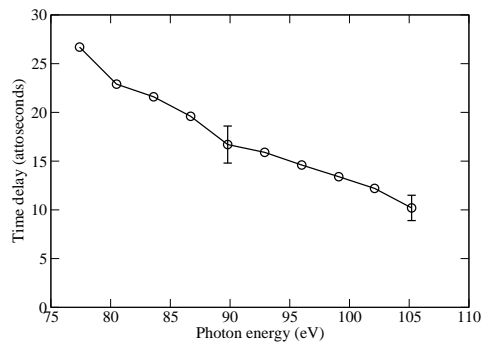}
\caption{\label{fig:delayvsenergy} Time delay in photoemission as a function of XUV photon energy.  The error bars shown on the two of the data points are $\pm$ one standard deviation.  The time delays were calculated using a single configuration representation of the wavefunction.}%
\end{center}
\end{figure}

To establish the dependence of the time-delay difference on photon energy,
we have repeated the calculations at several XUV photon energies.
Fig.~\ref{fig:delayvsenergy} shows  
that as the XUV photon energy is decreased from 105.2~eV to 77.4~eV, the 
time delay in photoemission increases from 10.2~as to 26.7~as.

One of the strengths of RMT is its ability to describe multi-electron 
excitations that may occur within the parent atom and the residual ion. 
We have repeated the time delay calculation using a greatly enlarged basis set 
for Ne$^+$. This basis is capable of including all single and double 
excitations from 2s2p$^6$ and 2s$^2$2p$^5$ to a set of 
pseudo-orbitals $\overline{3{\rm s}}$, $\overline{3{\rm p}}$ and 
$\overline{3{\rm d}}$.  As a consequence, for Ne, all single, double and 
triple excitations from 2s$^2$2p$^6$ to $\overline{3{\rm s}}$, 
$\overline{3{\rm p}}$ and $\overline{3{\rm d}}$ are included.  
Details of these orbital functions can be found in \cite{taylor1975}.
The calculation uses an XUV photon energy of 89.8~eV (for reasons described
below).
%
%
At this energy the average value of the time 
delay is found to be 14.5~as with standard deviation 1.5~as.  This compares to a 
value of 16.7~as obtained using the single-configuration representation. 
The single-configuration results are shown in Fig.~\ref{fig:delayvsenergy}. 
The error bar gives the estimated standard deviation (1.9~as) at 89.8~eV.

We cannot at present perform the larger basis set calculation on the
full range of frequencies in Fig.~\ref{fig:delayvsenergy}. Attempts to
do this at 105.2~eV, for example, failed because strong pseudo-resonances
were encountered \cite{taylor1975}.  These pseudo-resonances represent
channels omitted from the configuration-interaction representation of the wavefunction,
and their effect is to introduce additional, potentially spurious,
structure into the momentum spectrum of the outgoing wavepacket.
The resulting uncertainty in the time delay measurement overwhelms
the measurement itself.

The data points obtained at 89.8~eV nevertheless provides us with a
preliminary estimate of uncertainties in the calculations that
arise from various truncations of the {\it R}-Matrix basis sets. We cannot 
claim that the observed decrease in time delay is significant, but it points
to the possibility that the time delay in photoemission is sensitive 
to atomic structure and to the various excitations that may be possible 
within that structure, and suggests that future exploration of this
possibility is warranted.

Finally, we review some of the difficulties that will inevitably be
encountered in the future as attempts are made to improve on the
present analysis of this problem. 
Exploration of atomic-structure effects, especially at 105.2~eV, is 
complicated by the large number of double continua and double 
Rydberg states that are available. These include not only double 
continua attached to the 2s$^2$2p$^4$ and 2s2p$^5$ thresholds of 
Ne$^{2+}$, but also double continua/Rydberg series attached to the 
2s$^2$2p$^3$3s/3p/3d thresholds of Ne$^{2+}$.  For example,
2s$^2$2p$^3$3d$^2$4s states could be excited by single-photon absorption from the
2s$^2$2p$^6$ ground state via an admixture of 2s$^2$2p$^4$3d$^2$.  Such excitations
can be interpreted as the excitation of one 2p electron with a simultaneous collective
excitation of the residual five 2p electrons.  An accurate description of the resonance
structure thus requires a detailed description of triple excitations.  
The number of Ne$^+$ states that need to be included as target states to 
describe 2s$^2$2p$^3$3d$^2$4s resonances exceeds 150, even if only
excitations to 3s, 3p and 3d are included.  
A time-dependent calculation with this number of target states is not feasible
at present using our numerical methods, even on the largest available
massively parallel computers.  The inclusion of double continua
is also non-trivial, and, to the best of our knowledge, no group 
has attempted such calculations for the full Ne atom. 
Experiments at lower photon energy, around 75 eV,  would therefore be very useful, as 
this will greatly reduce the complexity of theoretical calculations.

In summary, we have applied the RMT method to the investigation of the time delay in photoemission from the 2s and 2p sub-shells of neon following absorption of an attosecond XUV pulse.  In order to reduce systematic differences between experiment and theory, we have used a similar approach as experiment in the analysis of our results.  At an XUV photon energy of 105.2~eV, we calculate a value for the time delay of $10.2 \pm 1.3$~as.  To date, all theoretical calculations of this time delay fall short of the experimental measurement of $21 \pm 5$~as. 
The RMT method
 yields a value for the time delay that is somewhat larger than the other theoretical calculations and that lies within two to three standard deviations of the experimental measurement.



We find that extraction of the time delay is extremely sensitive to even small features in the momentum spectra which can impact on the fitting processes.  However, analysis of the experimental data must also be subject to these issues.  To extract a value for the time delay, the experimental data are analyzed using a frequency-resolved optical gating (FROG) phase retrieval algorithm \cite{gagnon2008} which has been specially tailored for attosecond measurements.  It does interpolate the spectrogram along the energy axis but avoids interpolating along the delay axis. To enable a better comparison with experiment, it may be interesting for future theoretical work to generate full spectrograms which could then be subsequently analyzed using the same FROG phase retrieval algorithm.

\begin{acknowledgments}
LRM, HWvdH and KTT acknowledge funding from the UK Engineering and Physical Sciences Research Council. MAL and JSP acknowledge funding under the HECToR distributed CSE programme, which is provided through The Numerical Algorithms Group (NAG) Ltd.
\end{acknowledgments}

\bibliography{moorelPRA}

\end{document}